\documentclass[11pt,fleqn,twoside]{article}
\usepackage{amsfonts,latexsym}

\makeatletter
\newcommand{\prava}{\footnotesize\it
\begin{flushright}
\begin{minipage}{18cm}%{6cm}%9.6
Copyright \copyright 1998 by Olga Fushchych
\end{minipage}
\end{flushright}}

\newcommand{\name}[1]{\begin{flushleft}
                       \LARGE \bf #1
                       \end{flushleft}\vspace{-3mm}}

\newcommand{\Author}[1]{\begin{flushleft}
                       \it #1 \end{flushleft}}

\newcommand{\Adress}[1]{\begin{flushleft}
                       \it #1 \end{flushleft}}

\newcommand{\Date}[1]{\begin{flushleft}
                      \small  \it #1 \end{flushleft}}

\newcommand{\ehkol}{Author \ name}
\newcommand{\ohkol}{Article \ name}
\renewcommand{\@evenhead}{
\hspace*{-3pt}\raisebox{-15pt}[\headheight][0pt]{\vbox{\hbox to \textwidth
{\thepage \hfil \ehkol}\vskip4pt \hrule}}}
\renewcommand{\@oddhead}{
\hspace*{-3pt}\raisebox{-15pt}[\headheight][0pt]{\vbox{\hbox to \textwidth
{\ohkol \hfil \thepage}\vskip4pt\hrule}}}
\renewcommand{\@evenfoot}{}
\renewcommand{\@oddfoot}{}

\newcommand{\be}{\begin{equation}}
\newcommand{\ee}{\end{equation}}
\newcommand{\ba}{\hspace*{-5pt}\begin{array}}
\newcommand{\ea}{\end{array}}
\newcommand{\p}{\partial}
\newcommand{\ds}{\displaystyle}

\makeatother
\begin{document}

\setcounter{page}{159}

\thispagestyle{empty}

\newcommand{\pde}[2]{\frac{\p #1}{\p #2}} 
\newcommand{\ppd}[3]{\frac{\p^2 #1}{\p #2\p #3}} 
\newcommand{\pdd}[2]{\frac{\p^2 #1}{\p #2^2}} 

\renewcommand{\ehkol}{W. Fushchych}

\renewcommand{\ohkol}{Velocity of the Electromagnetic Field}

\begin{flushleft}
\footnotesize {\sf
Journal of Nonlinear Mathematical Physics} \qquad
{\sf  1998, V.5, N~2},\ 
\pageref{fushchych-fp}--\pageref{fushchych-lp}.\hfill {{\sc Article}}
\end{flushleft}

\vspace{-5mm}

{\renewcommand{\footnoterule}{}
{\renewcommand{\thefootnote}{}
\footnote{\prava}}

\name{What is the Velocity of the\\ Electromagnetic Field?}\label{fushchych-fp}

\Author{\fbox{Wilhelm FUSHCHYCH}}

\Adress{Institute of Mathematics, National Academy of Sciences of
Ukraine,\\
3 Tereshchenkivs'ka Str., 252004 Kyiv, Ukraine}

\Date{Translated by Marianna Euler and reprinted from Dopovidi
of the Academy of Sciences of Ukraine N4, p.51-53 (1997)}

\begin{abstract}
\noindent
A new definition for the electromagnetic field velocity is
proposed. The velocity depends on the physical fields.
\end{abstract}

%\label{fushchych-lp}

%\end{document}

%\subsection*{1. Introduction}

\strut\hfill

\noindent
The question posed by the title of this paper is, surprisingly, not
yet answered uniquely today; not even by way of definition.
According to modern assumptions the light is the electromagnetic field
(with corresponding frequencies) and therefore it is obvious that the
answer to the posed fundamental question is not obvious.

Today the follwing definitions of the velocity of light are used 
[1,2]: \newline
1) Phase velocity,\newline
2) Group velocity,\newline
3) Velocity of energy transport.

%\noindent
The definition of phase- and group velocity is based on assumptions
that the electromagnetic wave can be characterized by the function
$\Psi(t,\vec{x})$, which has the following form [1,2]
\be
\label{fushchych:eq1}
\Psi(t,\vec{x})=A(\vec{x})\cos(\omega t-g(\vec{x}))
\ee
or
\be
\label{fushchych:2}
\Psi(t,\vec{x})=\int_{0}^\infty A_\omega (\vec{x})\cos(\omega t-
g_\omega(\vec{x}))d\omega,
\ee
where $A(\vec x)$ is the wave amplitude and $g(\vec x)$ is an
arbitrary real function. The phase-velocity is defined by the
following formula
\be
\label{fushchych:eq3}
v_1=\frac{\omega}{|\vec{\bigtriangledown} g(\vec{x})|}
\ee
By the above formulas it is clear that the definition of the phase-
and group-velocity is based on the assumption that the electromagnetic
wave has the structure (1) (or (2)) and its velocity does not depend
on the amplitude $A$. Moreover, the 
equation which is to be satisfied by $\Psi$, has never been clearly
stated. This is, in fact, a very important point since $\Psi$ can
satisfy the standard linear wave equation (d'Alembert equation)
or, for example a nonlinear wave equation [3]. These two cases are
essentially different and lead to principly different results.
One should mention that the phase- and group-velocities cannot
directly be defined in terms of the electromagnetic fields $\vec E$
and $\vec H$.

%\medskip

The velocity of electromagnetic energy transport is defined by the
formula
\be
\vec v_2=\frac{\vec{S}}{W},\qquad \vec{S}=c(\vec{E}\times
\vec{H}),\qquad W=\vec{E}^2+\vec{H}^2,
\ee
where $\vec{S}$ is the Poyting-Heaviside vector.

%\medskip
\noindent
Formula (4) has the following disadvantage: Both $E$ and $H$ are invariant
under the Lorentz transformation, whereas $v_2$ does not have this
property.

%\medskip

The aim of the present paper is to give some new definitions of
the electromagnetic field velocity.

%\medskip

If the electromagnetic field is some enery flow, then we define
the velocity of
such flow, in analogy with hydrodynamics [4], by the following
equation
\be
\ba{l}
\ds \pde{\vec{v}}{t}+v_l\pde{\vec{v}}{x_l}=a_1(\vec{D},\vec{B}^2,
\vec{E}^2,\vec{H}^2, \vec{D}\cdot\vec{E},\ldots)\vec{D}
+a_2(\vec{D},\vec{B}^2,
\vec{E}^2,\vec{H}^2, \vec{D}\cdot\vec{E},\ldots)\vec{B}\\[2mm]
\qquad\ds +a_3(\vec{D},\vec{B}^2,
\vec{E}^2,\vec{H}^2, \vec{D}\cdot\vec{E},\ldots)\vec{E}
+a_4(\vec{D},\vec{B}^2,
\vec{E}^2,\vec{H}^2, \vec{D}\cdot\vec{E},\ldots)\vec{H}\\[2mm]
\qquad\ds +a_5(\vec{D},\vec{B}^2,
\vec{E}^2,\vec{H}^2, \vec{D}\cdot\vec{E},\ldots)\left(c(\vec\bigtriangledown
\times \vec H)-\vec{D}-4\pi \vec{J}\right)\\[2mm]
\qquad\ds +a_6(\vec{D},\vec{B}^2,
\vec{E}^2,\vec{H}^2, \vec{D}\cdot\vec{E},\ldots)\left(c(\vec\bigtriangledown
\times \vec H)+\vec{B}\right).
\ea
\ee
The structure and explicit form of the coefficients $a_1, \ldots, a_6$
is defined by the demand that equation (5) should be invariant
with
respect to the
Poincar\'e group if the fields are transformed according to the
Lorentz transformation [5].

%\medskip

The main advantage of (5), in comparison with (1), (2), lies in the
following:
\begin{enumerate}
\item The velocity of the electromagnetic field is directly
defined by the observables $\vec D,\vec B, \vec E,$ $\vec H,\vec J$, and
their first derivatives.
\item For particular coefficients, eq.(5) is invariant under the
Poincar\'e group.
\item In the case where $a_1=a_2=a_3=a_4=0$ and the fields $\vec D,
\vec B,\vec E,\vec H$ satisfy Maxwell's equation
\be
c(\vec \bigtriangledown\times \vec H)-\vec D-4\pi\vec J=0,\quad
c(\vec \bigtriangledown\times \vec E)+\vec B=0,
\ee
then the velocity of the electromagnetic field 
is of constant value, with 
\be
\pde{\vec{v}}{t}+v_l\pde{\vec{v}}{x_l}=0.
\ee
\end{enumerate}

In order to use eq.(5) one should concretely define the
coefficients $a_1,\ldots,a_6$.

\medskip

The explicitly-covariant definition of electromagnetic field velocity
can be given the following equation [5]
\be
v_\mu\pde{v_\alpha}{x^\mu}=a(\vec E^2,\vec H^2,\vec E\cdot\vec
H)F_{\alpha\beta}v^{\beta}.
\ee
Using Maxwell's equation in vacuum, one can obtain the following
formula for the velocity of the electromagnetic field
\be
|\vec v|=\left\{\frac{1}{2}\frac{(\p\vec E/\p t)^2+(\p\vec H/\p t)^2}
{(\mbox{rot}\ \vec E)^2 (\mbox{rot}\ \vec H)^2}\right\}^{1/2}
\ee
From (7) it is clear that the velocity depends only on derivatives
of the fields. $| \vec v|$ is a conditional invariant with respect to
the Lorentz transformation, i.e, if $\vec E$ and $\vec H$ satisfy
the full system of Maxwell's equations in vacuum, then $|\vec v|$ would
be an invariant of the Lorentz group. In other words, the conditional
invariant is a particular scalar combination of the fields,
for which the fields satisfy some equations with
nontrivial solutions. Well known invariants for the electromagnetic
field $\vec E\cdot\vec H$ and $\vec E^2-\vec H^2$ are absolute
invariants with respect to the Lorentz group.

\label{fushchych-lp}
\end{document}

\label{fushchych-lp}

\end{document}

\be \label{symenoh:symop1}
\ba{l}
\ds J_{ab}=x_{a} \p _{x_b} - x_{b} \p _{x_a}, \\[2mm]
\ds  Q_a= U _{a} \p _{{x}_{a}} + \frac{ i}{ 2}
\dot U _{a} x_{a}
\left(\psi \p_{ \psi}-\psi ^{\ast} \p_{ \psi ^{\ast}}\right)
 + \frac{ 1}{ 2} \ddot U_{a} x_{a} \p_{W},  \\[2mm]
\ds Q_{A}=2A \p_{t}+ \dot A x_{c} \p _{{x}_{c}}+
\frac{ i}{ 4} \ddot A x_{c} x_{c} \left(\psi \p_{\psi}
-\psi ^{\ast} \p_{ \psi ^{\ast}}\right) \\[2mm]
\ds \phantom{Q_{A}=} -\frac{ n\dot A}{ 2}\left(\psi \p_{\psi}
+\psi ^{\ast} \p_{ \psi ^{\ast}}\right)
+\left(\frac{ 1}{ 4} \stackrel{...}{A} x_{c}x_{c} -
2W \dot A \right) \p_{W}, \\[2mm]
\ds  Q_{B}= iB \left(\psi \p_{ \psi}-\psi ^{\ast} \p_{ \psi ^{\ast}}\right)
 + \dot B \p_{W}, \quad Z_1=\psi\p_{\psi},\quad Z_2=\psi^*\p_{\psi^*},
\ea \ee
Let us consider the following generalization of
 the Schr\"odinger equation

\be
\label{symenoh:eqSchrod}
i \frac{ \p\psi}{ \p t} +\Delta \psi +W(t, \vec{x}, |\psi|) \psi
+V_a(t,\vec x\,) \frac{ \p\psi }{ \p x_a} =0 ,
\ee
where
$\ds  \Delta= \frac{\p^2}{ \p x_a \p x_a}$, $a=\overline{1,n}$,
$\psi=\psi(t,\vec x\,)$ is an unknown complex function,
$ W=W(t,\vec x,|\psi|) $ and $V_a=V_a(t, \vec x\,)$ are potentials of
interaction.

When $V_a=0$ in (1), the standard Schr\"odinger equation is obtained.
Symmetry properties of this equation were thoroughly investigated (see,
e.g., \cite{symenoh:FSS}--\cite{symenoh:Tr}).
For arbitrary $ W(t, \vec{x}\,)$, equation (\ref{symenoh:eqSchrod})
admits only the trivial group of identical transformations
$\vec{x}\to {\vec{x}\,}'= \vec{x}$, $t\to t'=t$, $\psi \to \psi '= \psi$
\cite{symenoh:FSS,symenoh:Boy}.

In \cite{symenoh:F1}--\cite{symenoh:F3},  a method for extending
the symmetry group of equation (1) was suggested.
The idea lies in the fact that, in equation
(1), we assume that $ W(t, \vec{x}, |\psi|) $ is a new dependent variable
on equal conditions with $\psi$. This means that equation (1) is regarded
as a nonlinear equation even in the case where the potential $W$ does not
depend on $\psi$. Indeed, equation (1) is a set of
equations when $V$ is a certain set of arbitrary smooth functions.

\subsection*{2. Symmetry of the Schr\"odinger Equation with Potential}

Using this idea, we obtain the invariance algebra of the Schr\"odinger
 equation with potential, i.e.,
\be \label{symenoh:eqSchrod1}
i \frac{ \p\psi}{ \p t} +\Delta \psi +W(t, \vec{x}, |\psi|) \psi=0.
\ee

\noindent
{\bf Theorem 1.} {\it
Equation (2) is invariant under the infinite-dimensional
Lie algebra with infinitesimal operators of the form
\be \label{symenoh:symop1}
\ba{l}
\ds J_{ab}=x_{a} \p _{x_b} - x_{b} \p _{x_a}, \\[2mm]
\ds  Q_a= U _{a} \p _{{x}_{a}} + \frac{ i}{ 2}
\dot U _{a} x_{a}
\left(\psi \p_{ \psi}-\psi ^{\ast} \p_{ \psi ^{\ast}}\right)
 + \frac{ 1}{ 2} \ddot U_{a} x_{a} \p_{W},  \\[2mm]
\ds Q_{A}=2A \p_{t}+ \dot A x_{c} \p _{{x}_{c}}+
\frac{ i}{ 4} \ddot A x_{c} x_{c} \left(\psi \p_{\psi}
-\psi ^{\ast} \p_{ \psi ^{\ast}}\right) \\[2mm]
\ds \phantom{Q_{A}=} -\frac{ n\dot A}{ 2}\left(\psi \p_{\psi}
+\psi ^{\ast} \p_{ \psi ^{\ast}}\right)
+\left(\frac{ 1}{ 4} \stackrel{...}{A} x_{c}x_{c} -
2W \dot A \right) \p_{W}, \\[2mm]
\ds  Q_{B}= iB \left(\psi \p_{ \psi}-\psi ^{\ast} \p_{ \psi ^{\ast}}\right)
 + \dot B \p_{W}, \quad Z_1=\psi\p_{\psi},\quad Z_2=\psi^*\p_{\psi^*},
\ea \ee
where  $ U_{a}(t),A(t),B(t) $
are arbitrary smooth functions of $t$, over the index
$ c $ we mean summation from $1$ to $n$, $a,b=\overline{1,n}$, and over the
repeated index $a$ there is no summation. The upper dot stands for the
derivative with respect to time.}
}

\medskip

Note that the invariance algebra (\ref{symenoh:symop1}) includes the operators of
space ($U_a=1$) and time ($A=1/2$) translations, the Galilei operator
($U_a=t$), the dilation ($A=t$) and projective ($A=t^2/2$) operators.

\medskip

\noindent
{\bf Proof of Theorem 1.}
We seek the symmetry operators of equation
(\ref{symenoh:eqSchrod1}) in the class of f\/irst-order dif\/ferential operators of
the form:
\be \label{symenoh:oper1}
 X=\xi ^{\mu}(t,\vec x, \psi ,\psi ^{\ast}) \p _{x_{\mu}}+
\eta(t,\vec x, \psi ,\psi ^{\ast}) \p _{\psi }+
\eta ^{\ast}(t,\vec x, \psi ,\psi ^{\ast}) \p _{\psi ^{\ast}}+
\rho (t,\vec x,\psi , \psi ^{\ast},W)\p _W.
\ee
Using the invariance condition \cite{symenoh:FSS,symenoh:Ovs,symenoh:Olv}
 of equation (2) under  operator
(\ref{symenoh:oper1}) and the fact that  $ W=W(t,\vec x,|\psi|) $, i.e.,
$\ds \psi\frac{ \p W}{ \p\psi}=\psi^*\frac{\p W}{\p\psi^*}$,
we obtain the system of determining equations:
\be \label{symenoh:syst1}
\ba{l}
\ds \xi ^j_{\psi}=\xi ^j_{\psi^*}=0,\quad  \xi ^0_a=0,\  \xi ^a_a
=\xi ^b_b,\quad  \xi ^a_b+\xi ^b_a=0,\quad \xi ^0_0=2\xi ^a_a,\\[1mm]
\ds \eta_{\psi^*}=0,\quad \eta _{\psi \psi}=0,\quad \eta _{\psi a}
=(i/2) \xi ^a_0, \\[1mm]  \eta^*_{\psi}=0,\quad \eta^* _{\psi^* \psi^*}
=0,\quad \eta^*_{\psi^* a}=-(i/2) \xi ^a_0,\\[1mm]
\ds i\eta _0 +\eta _{cc}-\eta _{\psi}W\psi +2W\xi ^n_n \psi +
W\eta +\rho \psi =0,\\[1mm]
\ds -i\eta^* _0 +\eta^* _{cc}-\eta^* _{\psi^*}W\psi^* +2W\xi ^n_n \psi^* +
W\eta^* +\rho \psi^* =0,\\[1mm]
\ds \rho_{\psi}=\rho_{\psi^*}=0,
\ea \ee
where an index $j$ varies from 0 to $n$, $a,b=\overline{1,n}$,
over the repeated index $c$
we mean the summation from $1$ to $n$, and over the indices
$a,b$ there is no summation.

We solve system (\ref{symenoh:syst1}) and obtain the following result:
\[ \ba{l}
\ds \xi ^0=2A,\quad \xi ^a=\dot A x_a+ C^{ab}x_b +U _a,\quad a=\overline{1,n},
\\[2mm]
\ds \eta= \frac i2\left( \frac 12 \ddot A x_cx_c+\dot U _cx_c+B\right)\psi,
\quad
\ds \eta^*=-\frac i2 \left( \frac 12\ddot A x_cx_c+\dot U _cx_c+
E\right)\psi^* ,\\[4mm]
\ds \rho =\frac 12 \left( \frac 12 \stackrel{...}{ A} x_cx_c+
\ddot U _cx_c+\dot B\right)- \frac n2 i\ddot A-2W\dot A,
\ea \]
where $A,U_a,B$ are arbitrary functions of $t$, $E=B-2in\dot A+C_1$,
 $C^{ab}=-C^{ba}$ and $C_1$ are arbitrary constants.
The theorem is proved.

\medskip

The operators $ Q_B$ generate the f\/inite transformations:
\be \label{symenoh:tr1}
 \left\{ \ba{l}
t'=t,\quad  {\vec{x}\,}'=\vec{x}, \\[1mm]
\psi '=\psi \exp (iB(t)\alpha), \\[1mm]
\ds \psi^{*'}=\psi^* \exp (-iB(t)\alpha), \\[1mm]
W'=W+\dot{B}(t)\alpha,
\ea \right. \ee
where $\alpha$ is a group parameter, $B(t)$ is an arbitrary smooth
function.

Using the Lie equations, we obtain that the following transformations
correspond to the operators $Q_a$:
\be \label{symenoh:tr2}
  \left\{ \ba{l}
t'=t,\quad  x_a'=U _a(t)\beta _a+ x_a, \quad x_b'=x_b \ \ (b \not= a),\\[2mm]
\ds \psi '=\psi \exp \left(\frac{ i}{ 4}\dot U _aU _a\beta _a^2+
\frac{ i}{ 2}\dot U _ax_a\beta _a\right), \\[2mm]
\ds \psi^{*'}=\psi^* \exp \left(-\frac{ i}{ 4}\dot U _aU _a\beta _a^2-
\frac{ i}{ 2}\dot U _ax_a\beta _a\right), \\[2mm]
\ds W'=W+\frac{ 1}{ 2}\ddot U _a x_a\beta _a +\frac{ 1}{ 4}
\ddot U_a U _a \beta _a^2,
\ea \right. \ee
where $\beta _a (a=\overline{1,n}\,)$ are group parameters, $U_a=U_a(t)$
are arbitrary smooth functions,
there is no summation over the index $a$.
In particular, if $U _a (t)=t$, then the operators $Q_a$ are
the standard Galilei operators
\be \label{symenoh:opGal}
G_a=t\p _{x_a} + \frac{i}{2}x_a \left(\psi \p _{\psi}-\psi^*
 \p _{\psi^*}\right) .
\ee

For the operators $Q_A$, it is dif\/f\/icult to write out the f\/inite
transformations in the general form. We consider several particular cases:

\noindent
{\bf (a)} $A(t)=t$.
Then
\[
 Q_A=2t\p _t + x_c \p _{x_c}-\frac{ n}{ 2}
(\psi\p_{\psi}+\psi^*\p_{\psi^*}) -2W\p _W
\]
is a dilation operator
generating the transformations
\be \label{symenoh:tr3}
 \left\{ \ba{l}
t'=t\exp(2\lambda),\quad  x_c'=x_c\exp(\lambda),\\[2mm]
\ds \psi'=\exp\left(-\frac{ n}{ 2}\lambda\right)\psi,\quad \psi^{*'}
=\exp\left(-\frac{ n}{ 2}\lambda\right)\psi^*,\\[2mm]
W'=W\exp(-2\lambda),
\ea \right. \ee
where $\lambda$ is a group parameter.

\noindent
{\bf (b)} $A(t)=t^2/2$.
Then
\[
 Q_A=t^2\p _t + tx_c\p _{x_c} + \frac{ i}{ 4}x_cx_c\left(\psi \p
_{\psi}-\psi^*\p_{\psi^*}\right)-\frac{ n}{ 2}t\left(\psi\p_{\psi}+
\psi^*\p_{\psi^*}\right)  -2tW\p _W
\]
is the operator of projective transformations:
\be \label{symenoh:tr4}
 \left\{ \ba{l}
\ds t'=\frac{ t}{ 1-\mu t},\quad
 x_c'=\frac{ x_c}{ 1-\mu t},\\[3mm]
\ds \psi'=\psi(1-\mu t)^{n/2}\exp\left(\frac{ ix_cx_c\mu}{ 4(1-\mu t)}
\right),\\[3mm]
\ds  \psi^{*'}=\psi^*(1-\mu t)^{n/2}
\exp\left(\frac{ -ix_cx_c\mu}{ 4(1-\mu t)}\right),\\[2mm]
\ds W'=W(1-\mu t)^2,
\ea \right. \ee
$\mu$ is an arbitrary parameter.

Consider the example. Let
\be \label{symenoh:ex}
W=\frac{ 1}{{\vec x\,}^2}=\frac{ 1}{x_c x_c}.
\ee
We describe how new potentials are generated from potential
(\ref{symenoh:ex}) under transformations (\ref{symenoh:tr1}), (\ref{symenoh:tr2}), (\ref{symenoh:tr3}),
(\ref{symenoh:tr4}). \\
(i) $Q_B$:
\[
W= \frac{ 1}{ x_c x_c}\to W'=\frac{ 1}{ x_c x_c}+B(t)\alpha
\to
W''=\frac{ 1}{ x_c x_c}+B(t)(\alpha +\alpha ')\to\cdots,
\]
where $B(t)$ is an arbitrary smooth function, $\alpha$ and $\alpha'$ are
arbitrary real parameters.\\
(ii) $Q_a$:
\[ \ba{l}
\ds W=\frac{ 1}{ x_c x_c} \to W',\\[2mm]
\ds W'=\frac{ 1}{ (x_a-U _a(t)\beta _a)^2+ x_b
x_b}+\frac{ 1}{ 4}\ddot U_a U_a \beta _a^2+\frac{ 1}{ 2}\ddot
 U_a
\beta _a(x_a-U_a \beta _a),
\ea
\]
\[
\ba{l}
\ds W' \to W'',\\[2mm]
\ds W''=\frac{ 1}{ (x_a-U _a(t)(\beta _a +\beta _a'))^2+ x_b
x_b}+\frac{ 1}{ 4}\ddot U_a U_a (\beta _a^2 +\beta _a '^2)\\[4mm]
\ds \phantom{W''=}+\frac{ 1}{ 2}\ddot U_a (\beta _a+\beta _a ')
(x_a-U _a (\beta _a+\beta _a '))+\frac{ 1}{ 2}\ddot U_a U_a
\beta _a\beta _a '\to\cdots,
\ea
\]
where $U_a$ are arbitrary smooth functions, $\beta_a$ and $\beta_a'$ are
real
parameters, there is no summation over $a$ but there is summation over $b$
($b\not=a$).
In particular, if $U_a(t)=t$, then we have the standard Galilei operator
(\ref{symenoh:opGal}) and
\[
W=\frac{ 1}{ x_c x_c} \to W'=\frac{ 1}{ (x_a-
t\beta _a)^2+ x_b x_b}\to
W''=\frac{ 1}{ (x_a-t(\beta _a +\beta _a'))^2+ x_b x_b}\to\cdots
\]
(iii) $Q_A$ for $A(t)=t$ or $A(t)=t^2/2$ do not change the potential,
 i.e.,
\[
W=\frac{ 1}{ x_c x_c} \to W'=\frac{ 1}{ x_c x_c} \to
W''=\frac{ 1}{ x_c x_c}\to\cdots  \]

\subsection*{3. The Schr\"odinger Equation and Conditions for the Potential}

Consider several examples of the systems in which one of the equations is
equation (2) with potential $W=W(t,\vec x\,)$, and
the second equations is a certain condition for the potential~$W$. We f\/ind
the invariance algebras of these systems in the class of operators
\[  \ba{l}
\ds  X=\xi ^{\mu}(t, \vec x, \psi,\psi^* ,W) \p _{x_{\mu}}+
\eta (t, \vec x, \psi,\psi^*, W) \p _{\psi }\\[2mm]
\ds \phantom{X=}+\eta^* (t, \vec x, \psi,\psi^*, W) \p _{\psi^* }
+\rho (t, \vec x,\psi,\psi^* ,W)\p _W.
\ea \]
%**************************************************************************
(i) Consider equation (2) with the additional condition
for the potential, namely the Laplace equation.
\be \label{symenoh:system5} \left\{
\ba{l}
\ds i \frac{ \p\psi}{ \p t} +\Delta \psi +W(t, \vec{x}\,) \psi=0,\\[3mm]
\ds \Delta W =0.
\ea \right. \ee
%************************************************************************
System (\ref{symenoh:system5}) admits the
inf\/inite-dimensional Lie algebra with the inf\/initesimal operators
\be \label{symenoh:symop3}
\ba{l}
P_{0}=\p_t,\quad P_{a}=\p_{x_a}, \quad J_{ab}=x_{a}\p_{x_b} - x_{b}\p_{x_a},
\\[2mm]
\ds  Q_a= U_a\p_{x_a} + \frac{ i}{ 2} \dot U_a x_a
\left(\psi \p_{ \psi} -\psi ^{\ast} \p_{ \psi ^{\ast}}\right)
+ \frac{ 1}{ 2} \ddot U_{a} x_{a} \p_{W}, \ \ a=\overline{1,n}, \\[2mm]
\ds D=x_c \p_{x_c} +2t \p _t-\frac{ n}{ 2}\left(\psi\p_{\psi}+
\psi^*\p_{\psi^*}\right) -2W\p _W, \\[2mm]
\ds  A=t^2 \p _t + t x_{c} \p _{{x}_{c}}+
\frac{ i}{ 4}  x_{c} x_{c} \left(\psi \p_{\psi}-
\psi ^{\ast} \p_{ \psi ^{\ast}}\right)-
\frac{ n}{ 2}t\left(\psi\p_{\psi}+
\psi^*\p_{\psi^*}\right) -2Wt  \p_{W}, \\[2mm]
\ds  Q_{B}= iB (\psi \p_{ \psi} -\psi ^{\ast} \p_{ \psi ^{\ast}})
+ \dot B \p_{W}, \quad Z_1=\psi \p _{\psi},\ Z_2=\psi^*\p_{\psi^*},
\ea \ee
where $U_a (t)\ (a=\overline{1,n}\,)$ and $B(t)$ are arbitrary
smooth functions. In particular, algebra~(\ref{symenoh:symop3}) includes the
Galilei operator (\ref{symenoh:opGal}).\\
%***************************************************************************
(ii) The condition for the potential is the heat equation.
\be \label{symenoh:system1}
 \left\{ \ba{l}
\ds i \frac{ \p\psi}{ \p t} +\Delta \psi +W(t, \vec{x}\,) \psi=0,\\[3mm]
W_0+\lambda \Delta W=0.
\ea \right. \ee

The maximal invariance algebra  of system (\ref{symenoh:system1}) is
\[\ba{l}
P_0=\p _t, \quad P_a=\p _{x_a}, \quad J_{ab}=x_{a}\p_{x_b} - x_{b}\p_{x_a},
\\[2mm]
\ds D=2t\p _t +x_c \p _{x_c}-\frac{ n}{ 2}\left(\psi\p_{\psi}+
\psi^*\p_{\psi^*}\right) -2W \p _W, \\[2mm]
\ds Z_1=\psi \p _{\psi}, \quad Z_2=\psi^*
\p _{\psi^*}, \quad Z_3=it \left(\psi \p _{\psi}-\psi^* \p _{\psi^*}
\right) +\p _W.
\ea \]
(iii) The condition for the potential is the wave equation.
\be \label{symenoh:system2}
  \left\{ \ba{l}
\ds i \frac{ \p\psi}{ \p t} +\Delta \psi +W(t, \vec{x}\,) \psi=0,\\[3mm]
\Box W=0.
\ea \right. \ee

The maximal invariance algebra of system (\ref{symenoh:system2}) is
\[\ba{l}
P_0=\p _t, \quad P_a=\p _{x_a}, \quad J_{ab}=x_{a}\p_{x_b} - x_{b}\p_{x_a},
 \quad
\ds Z_1=\psi \p _{\psi}, \quad Z_2=\psi^* \p _{\psi^*}, \\[2mm]
\ds Z_3=it \left(\psi \p _{\psi}-\psi^* \p _{\psi^*}\right) +\p _W,
\quad Z_4=
it^2 \left(\psi \p _{\psi}-\psi^* \p _{\psi^*}\right) +2t\p _W.
\ea \]
(iv) The condition for the potential is the Hamilton-Jacobi equation.
 \be \label{symenoh:system4}
 \left\{ \ba{l}
\ds i \frac{ \p\psi}{ \p t} +\Delta \psi +W(t, \vec{x}\,) \psi=0,\\[4mm]
\ds \frac{ \p W}{ \p t}- \lambda \frac{ \p W}{ \p x_a}
\frac{ \p W}{ \p x_a}=0.
\ea \right. \ee

The maximal invariance algebra is
\[
\ba{l}
\ds P_0=\p _t, \quad P_a=\p _{x_a},  \quad
 J_{ab}=x_{a}\p_{x_b} - x_{b}\p_{x_a}, \\[2mm]
Z_1=\psi \p _{\psi}, \quad Z_2=\psi^*\p _{\psi^*}, \quad
Z_3=it (\psi \p _{\psi}-\psi^* \p _{\psi^*}) +\p _W.
\ea \]
(v) Consider very important and interesting case in $(1+1)$-dimensional
space-time where
the condition for the potential is the KdV equation.
\be \label{symenoh:system3}
 \left\{  \ba{l}
\ds i \frac{ \p\psi}{ \p t} +\frac{ \p^2\psi}{ \p x^2}
+W(t,x) \psi=0,\\[3mm]
\ds \frac{ \p W}{ \p t} +\lambda _1 W \frac{ \p W}{ \p x}
+\lambda _2 \frac{ \p^3 W}{ \p x^3} =F(|\psi|), \ \ \lambda _1 \not= 0.
\ea \right. \ee

For an arbitrary $F(|\psi|)$, system (\ref{symenoh:system3}) is invariant
 under
the Galilei operator and the maximal invariance algebra is the following:
\be \label{symenoh:AI3}
\ba{l}
\ds P_0=\p_t, \quad  P_1=\p_x, \quad
Z=i\left(\psi \p_{\psi}-\psi^* \p_{\psi^*}\right), \\[2mm]
\ds G=t \p _x +\frac{ i}{ 2}\left(x+\frac{ 2}{ \lambda _1}t\right)
\left(\psi \p_{\psi}-\psi^*\p_{\psi^*}\right) +
\frac{ 1}{ \lambda _1}\p _W.
\ea \ee
For $F=C=\mbox{const}$, system
(\ref{symenoh:system3}) admits the extension, namely,  it
is invariant under the algebra $\langle P_0,P_1,G,Z_1,Z_2\rangle$,
where $P_0,P_1,G$
have the form (\ref{symenoh:AI3}) and $Z_1=\psi\p_{\psi}$,
$Z_2=\psi^*\p_{\psi^*}$.

The Galilei operator $G$ generates the following transformations:
\[
\left\{  \ba{l}
\ds t'=t,\quad x'=x+\theta t,\quad  W'=W+\frac{ 1}{ \lambda _1}\theta,\\[2mm]
\ds \psi '=\psi \exp \left(\frac{ i}{ 2}\theta x
+\frac{ i}{ \lambda _1}\theta t +\frac{ i}{ 4}
\theta ^2 t\right), \\[2mm]
\ds \psi^{*'}=\psi^* \exp \left(-\frac{ i}{ 2}\theta x
-\frac{ i}{ \lambda _1}\theta t -\frac{ i}{ 4}
\theta ^2 t\right),
\ea \right. \]
where $\theta$ is a group parameter. Here, it is important that
$\lambda_1\not=0$, since otherwise, system (\ref{symenoh:system3}) does not
admit the Galilei operator.

\subsection*{4. Finite-dimensional Subalgebras}

Algebra (\ref{symenoh:symop1}) is inf\/inite-dimensional. We se\-lect
cer\-tain f\/i\-nite-di\-men\-sio\-nal subalgebras from it.
In particular, we give the examples of functions
$U_a(t)$ and $B(t)$, for which the subalgebra generated by the operators
\be \label{symenoh:subalg}
P_0, \ P_a, \ J_{ab}, \ Q_a, \ Q_B, \ Z_1, \ Z_2
\ee
is f\/inite-dimensional.

\noindent
{\bf (a)} $U_a(t)=\exp(\gamma t)$.
In this case,  subalgebra (\ref{symenoh:subalg}) has the form
\[ \ba{l}
P_0, \ P_a, \ J_{ab},\ Z_1,\ Z_2, \\[2mm]
\ds Q_a=e^{\gamma t}\left(\p _{x_a}+\frac{ i}{ 2}
\gamma x_a \left(\psi \p _{\psi}-\psi^* \p _{\psi^*}\right)+
\frac{ 1}{ 2} \gamma ^2 x_a \p _W\right),\quad a=\overline{1,n}, \\[2mm]
\ds Q_B=e^{\gamma t}\left(i\psi \p _{\psi}-
i\psi^* \p _{\psi^*}+\gamma \p _W\right).
\ea \]
{\bf (b)}  $U _a(t)=C_1\cos (\nu t)+C_2 \sin (\nu t)$.
Then  subalgebra (\ref{symenoh:subalg}) has the form:
\[
\ba{l}
P_0, \ P_a,\ J_{ab},\ Z_1,\ Z_2,\\[2mm]
\ds  Q^{(1)}_a=\cos (\nu t)\p _{x_a}-\frac{ i}{ 2}\nu \sin (\nu
t)x_a \left(\psi \p _{\psi} -\psi^* \p _{\psi^*}\right)-
\frac{ 1}{ 2}\nu ^2 \cos (\nu t) x_a \p _W,
\\[2mm]
\ds  Q^{(2)}_a=\sin (\nu t)\p _{x_a}+\frac{ i}{ 2}\nu \cos (\nu
t)x_a \left(\psi \p _{\psi}-\psi^* \p _{\psi^*}\right) -
\frac{ 1}{ 2}\nu ^2 \sin (\nu t) x_a \p _W,
\\[2mm]
\ds  X_1=i\sin (\nu t)\left(\psi \p _{\psi}-\psi^* \p _{\psi^*}\right)
+\nu \cos (\nu t) \p _W,\\[2mm]
\ds X_2=i\cos (\nu t)\left(\psi \p _{\psi}-\psi^* \p _{\psi^*}\right)
-\nu \sin (\nu t) \p _W.
\ea
\]
{\bf (c)} $U_a (t)=C_1t^k+C_2t^{k-1}+\cdots+C_kt+C_{k+1}$.
Then  subalgebra (\ref{symenoh:subalg}) has the form:
\[ \ba{l}
P_0, \ P_a,\ J_{ab},\ Z_1, \ Z_2,\\[2mm]
\ds Q^{(1)}_a=t^k\p _{x_a}+\frac{ i}{ 2}kt^{k-1}x_a\left(\psi
\p _{\psi}-\psi^* \p _{\psi^*}\right)+
\frac{ 1}{ 2}k(k-1)t^{k-2}x_a\p _W, \\[2mm]
\ds Q^{(2)}_a=t^{k-1}\p _{x_a}+\frac{ i}{ 2}(k-1)t^{k-2}x_a\left(\psi
\p _{\psi}-\psi^* \p _{\psi^*}\right) +\frac{ 1}{ 2}(k-1)(k-2)t^{k-3}x_a\p _W, \\[2mm]
\cdots \\[2mm]
\ds Q^{(k)}_a=t\p _{x_a}+\frac{ i}{ 2}x_a
\left(\psi \p _{\psi}-\psi^* \p _{\psi^*}\right), \\[2mm]
\ds Q^{(1)}_B=it\left(\psi \p _{\psi}-\psi^* \p _{\psi^*}\right)+\p _W,\\[2mm]
\cdots\\[2mm]
\ds Q^{(2k-2)}_B=it^{2k-2}\left(\psi \p _{\psi}-
\psi^* \p _{\psi^*}\right)+(2k-2)t^{2k-3}\p _W.
\ea \]

\subsection*{5. The Schr\"odinger Equation with Convection Term}

Consider equation (1) for $W=0$, i.e., the Schr\"odinger equation with
convection term
\be \label{symenoh:conv1}
\ds i\frac{ \p\psi}{\p t}+\Delta\psi=V_a
\frac{ \p\psi}{\p x_a},
\ee
where $\psi$ and $V_a$ $(a=\overline{1,n}\,)$ are complex functions
of $t$ and $\vec x$. For extension of symmetry, we again regard the functions
$V_a$ as dependent variables. Note that the requirement that the functions
$V_a$ are complex is essential for symmetry of (\ref{symenoh:conv1}).

Let us investigate symmetry properties of
(\ref{symenoh:conv1}) in the class of f\/irst-order dif\/ferential operators
\[
 X=\xi ^{\mu}\p _{x_{\mu}}+\eta \p _{\psi }+
\eta ^{\ast}\p_{\psi ^{\ast}}+\rho^a \p _{V_a}+
\rho^{*a} \p _{V^*_a},
\]
where $\xi^{\mu},\eta,\eta^*,\rho^a,\rho^{*a}$ are functions of $t,\vec x,
\psi,\psi^*,V_a,V^*_a$.

\medskip

\noindent
{\bf Theorem 2.} {\it Equation (\ref{symenoh:conv1}) is invariant under the
infinite-dimensional Lie algebra with the infinitesimal operators
\be \label{symenoh:111}
\ba{l}
\ds Q_A=2A\p_t+\dot A x_c\p_{x_c}-i\ddot A x_c\left(\p_{V_c}-\p_{V^*_c}\right)
-\dot A
\left(V_c \p_{V_c}+V^*_c\p_{V^*_c}\right),\\[2mm]
\ds Q_{ab}=E_{ab}\left(x_a\p_{x_b}-x_b\p_{x_a}+V_a\p_{V_b}-V_b\p_{V_a}+
V^*_a\p_{V^*_b}-V^*_b\p_{V^*_a}\right)\\[2mm]
\phantom{Q_{ab}=}-i\dot E_{ab}\left(x_a\p_{V_b}-x_b\p_{V_a}-x_a\p_{V^*_b}
+x_b\p_{V^*_a}\right), \\[2mm]
\ds Q_a=U_a\p_{x_c}-i\dot U_a\left(\p_{V_a}-\p_{V^*_a}\right),\\[2mm]
\ds Z_1=\psi\p_{\psi},\quad Z_2=\psi^{\ast}\p_{\psi^{\ast}},\quad Z_3
=\p_{\psi},\quad Z_4=\p_{\psi^{\ast}},
\ea \ee
where $A,E_{ab},U_a$ are arbitrary smooth functions of $t$. We mean
summation over the index $c$ and no summation over indices $a$ and $b$.}

\medskip

This theorem is proved by analogy with the previous one.

Note that algebra (\ref{symenoh:111}) includes, as a particular case,
the Galilei
operator of the form:
\be \label{symenoh:222}
G_a=t\p_{x_a}-i\p_{V_a}+i\p_{V^*_a}.
\ee
This operator generates the following f\/inite transformations:
\[ \left\{ \ba{l}
t'=t, \quad x_a'=x_a+\beta_a t,\quad  x_b'=x_b\ (b\not=a),\\[1mm]
\psi'=\psi,\quad  \psi^{*'}=\psi^*,\\[1mm]
V_a'=V_a-i\beta_a,\quad  V^{*'}_a=V^*_a+i\beta_a,
\ea \right.
\]
where $\beta_a$ is an arbitrary real parameter.
Operator (\ref{symenoh:222}) is essentially dif\/ferent from the standard Galilei
opera\-tor (\ref{symenoh:opGal}) of the Schr\"o\-din\-ger equa\-tion, and we cannot
derive operator~(\ref{symenoh:opGal}) from algebra (\ref{symenoh:111}).

Consider now the system of equation (\ref{symenoh:conv1}) with the additional
condition for the potentials $V_a$, namely, the complex Euler equation:
\be \label{symenoh:333} \left\{
\ba{l}
\ds i\frac{ \p\psi}{\p t}+\Delta\psi=V_a
\frac{ \p\psi}{\p x_a},\\[4mm]
\ds i\frac{ \p V_a}{\p t}-V_b\frac{ \p V_a}{\p x_b}=
F(|\psi|)\frac{ \p\psi}{\p x_a}.
\ea \right. \ee
Here, $\psi$ and $V_a$ are complex dependent variables of $t$ and $\vec x$,
$F$ is an arbitrary function of $|\psi|$. The coef\/f\/icients of the second
equation of the system provide the broad symmetry of this system.

Let us investigate the symmetry classif\/ication of system (\ref{symenoh:333}).
Consider the following f\/ive cases.\\
{\bf 1.} $F$ is an arbitrary smooth function.
 The maximal invariance algebra is $\langle P_0,P_a,J_{ab},G_a\rangle$, where
\[ \ba{l}
\ds P_0=\p_t, \quad  P_a=\p_{x_a},\\[2mm]
\ds J_{ab}=x_a\p_{x_b}-x_b\p_{x_a}+V_a\p_{V_b}-V_b\p_{V_a}+
V^{\ast}_a\p_{V^{\ast}_b}-V^{\ast}_b\p_{V^{\ast}_a},\\[2mm]
\ds G_a=t\p_{x_a}-i\p_{V_a}+i\p_{V^{\ast}_a}.
\ea \]
{\bf 2.} $F=C|\psi|^k$, where $C$ is an arbitrary complex constant, $C\not=0$,
 $k$ is an arbitrary real number, $k\not=0$ and $k\not=-1$.
 The maximal invariance algebra is $\langle P_0,P_a,J_{ab},G_a,D^{(1)}\rangle$, where
\[
D^{(1)}=2t\p_t+x_c\p_{x_c}-V_c \p_{V_c}-V^{\ast}_c\p_{V^{\ast}_c}-
\frac{ 2}{ 1+k}(\psi\p_{\psi}+\psi^*\p_{\psi^*}).
 \]
{\bf 3.} $\ds F=\frac{ C}{ |\psi|}$, where $C$ is an arbitrary complex constant,
$C\not=0$.
 The maximal invariance algebra is $\langle P_0,P_a,J_{ab},G_a,Z=Z_1+Z_2
 \rangle $, where
\[ Z=\psi\p_{\psi}+\psi^{\ast}\p_{\psi^{\ast}},\quad  Z_1=\psi
\p_{\psi},\quad  Z_2=\psi^*\p_{\psi^*}. \]
{\bf 4.} $F=C\not=0$, where $C$ is an arbitrary complex constant.
 The maximal invariance algebra is $\langle P_0,P_a,J_{ab},
G_a,D^{(1)},Z_3,Z_4\rangle $, where
\[ Z_3=\p_{\psi},\ Z_4=\p_{\psi^{\ast}}.\]
{\bf 5.} $F=0$.  The maximal invariance algebra is $\langle P_0,P_a,J_{ab},
G_a,D,A,Z_1,Z_2,Z_3,Z_4\rangle$, where
\[ \ba{l}
\ds D=2t\p_t+x_c\p_{x_c}-V_c \p_{V_c}-V^{\ast}_c\p_{V^{\ast}_c},\\[2mm]
\ds A=t^2\p_t+tx_c\p_{x_c}-(ix_c+tV_c)\p_{V_c}+(ix_c-tV^*_c)\p_{V^*_c}.
\ea \]

\subsection*{6. Contact Transformations}

Consider the two-dimensional Schr\"odinger equation
\be \label{symenoh:eq51}
i\psi _t +\psi _{xx}=V(t,x,\psi,\psi _x, \psi _t).
\ee

We seek the inf\/initesimal operators of contact transformations in the
class of the f\/irst-order dif\/ferential operators of the form
\cite{symenoh:FSS,symenoh:Olv}
\be \label{symenoh:symop51} \ba{l}
\ds X=\xi ^{\nu}(t,x,\psi, \psi _t, \psi _x)\p _{x_{\nu}}+
\eta (t,x,\psi, \psi _t, \psi _x)\p _{\psi}  \\[2mm]
\ds \phantom{X=}+\zeta ^{\nu}(t,x,\psi, \psi _t, \psi _x)\p _{\psi _{\nu}}+
\mu (t,x,\psi, \psi _t, \psi _x,V)\p _V,
\ea \ee
where
\be \label{symenoh:defin}
\xi ^{\nu}=-\frac{ \p W}{ \p \psi _{\nu}}, \quad
\eta=W-\psi _{\nu}\frac{ \p W}{ \p \psi _{\nu}}, \quad
\zeta ^{\nu}=\frac{ \p W}{ \p x_{\nu}}+
\psi _{\nu}\frac{ \p W}{ \p \psi }
\ee
for a function $W=W(t,x,\psi,\psi _x, \psi _t)$.
The condition of invariance of equation
(\ref{symenoh:eq51}) under operators
(\ref{symenoh:symop51}),
(\ref{symenoh:defin}) implies that the unknown function $W$ has
the form
\[
W=F^1(t)\psi _t +F^2(t,x,\psi, \psi _x),
\]
where $F^1$ and $F^2$ are arbitrary functions of their arguments.

Then
\[ \ba{l}
\xi ^0=-F^1(t), \quad \xi ^1=-F_{\psi _x}^2(t,x,\psi,\psi _x),\\[2mm]
\ds \eta =F^2-\psi _xF_{\psi _x}^2,\quad \zeta ^0=F_t^1 \psi _t
+F_t^2+\psi _tF_{\psi}^2,\quad  \zeta ^1=F_x^2+\psi _x F_{\psi}^2,\\[2mm]
\mu=i(W_t+\psi _tW_{\psi})+W_{xx}+2W_{x\psi}\psi _x\\[2mm]
\ds \phantom{\mu=}-(i\psi _t-V)
\left(W_{x\psi_x}+W_{\psi}+\psi _xW_{\psi \psi _x}\right)+
(\psi _x)^2W_{\psi \psi}\\[2mm]
\ds \phantom{\mu=}-(i\psi _t -V)\left(W_{x\psi _x}+\psi _x W_{\psi
\psi _x}-(i\psi _t -V)W_{\psi _x \psi _x}\right).
\ea \]
Thus, equation (\ref{symenoh:eq51}) is invariant under the inf\/inite-dimensional
group of contact transformations with the inf\/initesimal operators:
\[ \ba{l}
\ds Q_{F^1}=-F^1\p _t+F^1_t\psi_t\p _{\psi _t}+iF^1_t\psi_t\p_V, \\[2mm]
\ds Q_{F^2}=-F^2_{\psi_x}\p _x+(F^2-\psi _x F^2_{\psi _x})\p _{\psi}+
(F^2_t+\psi _t F^2_{\psi})\p _{\psi _t}\\[2mm]
\ds \phantom{Q_{F^2}=}+(F^2_x+\psi _x F^2_{\psi})
\p _{\psi _x}+\Bigl\{iF^2_t+i\psi _tF^2_{\psi}+F^2_{xx}+2F^2_{x\psi}\psi _x\\[2mm]
\ds \phantom{Q_{F^2}=}+(\psi _x)^2 F^2_{\psi \psi}-(i\psi _t-V)
(2F^2_{x \psi _x}+2\psi _x
F^2_{\psi \psi _x}+F^2_{\psi})+(i\psi _t-V)^2F^2_{\psi _x \psi
_x}\Bigr\}\p _V,
\ea \]
where $F^1=F^1(t)$ and $F^2=F^2(t,x,\psi,\psi _x)$ are arbitrary functions.

Consider the special case. Let $F^1(t)=1$, $F^2(t,x,\psi, \psi _x)=
-(\psi _x)^2.$ Then $W=\psi _t -(\psi _x)^2.$
The operators of the contact transformations have the form
\[ Q_{F^1}=\p _t,\]
\be \label{symenoh:contoper}
Q_{F^2}=2\psi _x\p _x+(\psi _x)^2\p _{\psi}-2(i\psi _t-V)^2\p _V.
\ee
The operator (\ref{symenoh:contoper}) generate the f\/inite transformations:
\be \label{symenoh:tr52} \left\{
\ba{l}
\ds x'=2\psi _x\theta +x,\quad  t'=t,\\[1mm]
\ds \psi '=(\psi _x)^2\theta+\psi,\quad \psi _x'=\psi _x, \quad
 \psi _t'=\psi _t,\\[2mm]
\ds V'=\frac{ 2i\theta (V-i\psi _t)\psi _t+V}{ 2\theta (V-i\psi _t)+1}.
\ea  \right. \ee
Transformations (\ref{symenoh:tr52}) can be used for generating
exact solutions of equation
(\ref{symenoh:eq51}) from the known solution and for constructing nonlocal
ansatzes reducing the given equation to the system of ordinary
dif\/ferential equations.

\label{symenoh-lp}


\begin{thebibliography}{5}
\footnotesize
\bibitem{1} Born M. and Wolf E., Principles of Optics, MacMillan,
New York.

\bibitem{2}Brillouin L., Wave Propagation and Group Velocity,
Academic press, New York, 1960.

\bibitem{3} Fushchych W., Shtelen W. and Serov N., Symmetry Analysis
and Exact Solution of Equations of Nonlinear Mathematical Physics.,
Kluwer Academic press, 1993.

\bibitem{4} Fushchych W., {\it Dopovidi of the Ukrainian Academy of
Sciences}, 1992, Nr.4, p.24.

\bibitem{5} Fushchych W., {\it J. Nonlin. Math. Phys.}, 1995, V.2,
p.216.

\end{thebibliography}

\begin{thebibliography}{15}
\footnotesize

\bibitem{symenoh:FSS} Fushchych W., Shtelen V. and Serov N.,  Symmetry
Analysis and Exact Solutions of Equations of Nonlinear Mathematical Physics,
Kluwer Academic Publishers, Dordrecht, 1993.
\bibitem{symenoh:FN} Fushchych W. and Nikitin A.,  Symmetry of Equations
of Quantum Mechanics, Allerton Press, New York, 1994.
\bibitem{symenoh:Boy} Boyer C., The maximal 'kinematical' invariance
group for an arbitrary potential, {\it Helv. Phys. Acta}, 1974, V.47,
589--605.
\bibitem{symenoh:Tr} Truax D.R., Symmetry of time-dependent Schr\"odinger
equations. I. A classif\/ication of time-dependent potentials by their
maximal kinematical algebras, {\it J. Math. Phys.}, 1981, V.22, N~9,
1959--1964.
\bibitem{symenoh:F1} Fushchych W., How to extend symmetry of dif\/ferential
equations?, in:  Symmetry and Solutions of Nonlinear Equations of
Mathematical Physics, Inst. of Math. Ukrainian
Acad. Sci., Kyiv, 1987, 4--16.
\bibitem{symenoh:F2} Fushchych W., New nonlinear equations for electromagnetic
f\/ield having the velocity dif\/ferent from~$c$, {\it Dopovidi Academii Nauk
Ukrainy}, 1992, N~4,  24--27.
\bibitem{symenoh:F3} Fushchych W., Ansatz '95, {\it J. Nonlin. Math. Phys.},
1995, V.2, N~3--4,  216--235.
\bibitem{symenoh:Ovs} Ovsyannikov L.V.,  Group Analysis of Dif\/ferential
Equations, Academic Press, New York, 1982.
\bibitem{symenoh:Olv} Olver P., Application of Lie Groups to
Dif\/ferential Equations, Springer, New York, 1986.
\end{thebibliography}
\end{document}